\documentclass[12pt]{article}
\usepackage{amsmath,amssymb}
\usepackage{graphicx}
\usepackage{epstopdf}
\usepackage[pdftex]{hyperref}
\DeclareGraphicsExtensions{.eps,.bmp,.wmf,.jpeg,.pdf}
\numberwithin{equation}{section}
\def\be{\begin{equation}}
\def\ee{\end{equation}}

\def\bea{\begin{eqnarray}}

\def\eea{\end{eqnarray}}
\title{Non-minimal kinetic coupling and the phenomenology of dark energy}
\author{L.N. Granda\thanks{ngranda@univalle.edu.co, ngranda@um.es} \\{\small\it Departamento de Fisica, Universidad del Valle}\\{\small\it A.A. 25360, Cali, Colombia}\\{\small\it and}\\
{\small\it Departamento de Fisica, Universidad de Murcia}\\{\small\it 30100 Murcia, Spain}}
\date{}
\begin{document}
\maketitle

\begin{abstract}
\noindent We study a model of scalar field with a general non-minimal kinetic coupling to itself and to the curvature. The cosmological dynamics of this model and the issue of accelerated expansion is analyzed. Solutions giving rise to power law expansion have been found. By constraining the potential of the model, we obtained a variety of solutions corresponding to phenomenologically acceptable models of dark energy, like the $\Lambda$CDM model, variable cosmological constant model VCC, FRW with two dimensional topological
defects, FRW with phantom dark energy, showing that this model with the appropriate potential in each case, can describe a rich variety of dynamical scenarios.
\end{abstract}

\section{Introduction}
\noindent 
A wide range of cosmological observations indicate
that the universe has entered a phase of accelerating expansion, which becomes one of the important
puzzles of the contemporary physics. Those observations include the type Ia supernovae (SnIa) standard candles \cite{hicken}, \cite{kowalski}, the angular location of the first peak in the CMB power spectrum \cite{komatsu} and baryon acoustic oscillations of the matter density power spectrum \cite{percival}. These evidences represent a great stimulus for theoretical work and originated the concept of Dark Energy. The Dark Energy (DE) models attribute the observed accelerating expansion to an unknown energy component with negative pressure, which dominates
the universe at recent cosmological times. The simplest model describing the cosmic expansion
rate with a high degree of accuracy, is the cosmological constant \cite{peebles}, \cite{padmana1} in a spatially flat space, which provides a constant equation of state. However this physically motivated model is plagued by fine tuning problems \cite{zlatev}. Alternatively,
dark energy may be described by different dynamical scalar field models with time-dependent equation of state, like quintessence \cite{RP}, \cite{wett}; string theory fundamental scalar known as tachyon \cite{pad}; K-essence models involving a generalized form of the kinetic energy \cite{stein3},\cite{chiba}; scalar field with negative kinetic term, which provides a solution known
as phantom dark energy \cite{caldwell} (see \cite{copeland} for a review). An alternative description of DE may be given by perfect fluids with adequate equation of state, like Chaplygin gas \cite{bilik}.
Another possibilities to the explanation of the DE are represented by the scalar-tensor theories, which contain a direct coupling of the scalar field to the curvature, providing in principle a mechanism to evade the coincidence problem, and naturally allowing (in some cases) the crossing of the phantom barrier \cite{perivo}. The non-minimal coupling between the quintessence field and curvature, as an explanation of the accelerated expansion, has been considered among others, in refs. \cite{chiba1} \cite{uzan},\cite{wands},\cite{easson1}. An inflationary model including the DBI term and non-minimally coupled scalar has been proposed in \cite{easson2,easson3}.\\
\noindent A model that can be regarded as falling within the scalar-tensor theories is the model with non-minimally coupled kinetic terms to the curvature. In this paper we consider an explicit coupling between the scalar field, the kinetic term and the curvature, as a source of dark energy, and analyze the role of this  coupling in an evolution scenario with late-time accelerated expansion \cite{granda, granda1}. The basic motivation for studying such theories is related with the fact that they appear as low energy limit of several higher dimensional theories, e.g. superstring theory \cite{green}, and provide a possible approach to quantum gravity from a perturbative point of view \cite{donoghue}. A coupling between curvature and kinetic terms also appears as part of the Weyl anomaly in $N=4$ conformal supergravity \cite{tseytlin, odintsov0}. A model with non-minimal derivative couplings was proposed in \cite{amendola2}, \cite{capozziello1}, \cite{capozziello2} in the context of inflationary cosmology, and recently, non-minimal derivative coupling of the Higgs field was considered in \cite{germani}, also as inflationary model. In \cite{caldwell1} a derivative coupling to Ricci tensor has been considered to study cosmological restrictions on the coupling parameter, and the role of this coupling during inflation. Some asymptotical solutions for a non-minimal kinetic coupling to scalar and Ricci curvatures were found in \cite{sushkov}, and quintessence and phantom cosmological scenarios with non-minimal derivative coupling have been studied in \cite{saridakis}. A scalar field with kinetic term coupled to a product of Einstein tensors has been considered in \cite{gao}, and a vector field non-minimally coupled to curvature has been studied in \cite{beltran}. A kinetic coupling of scalar field to Ricci curvature appears as part of the action in a class of covariant renormalizable gravity theories \cite{sergio1}.\\
The modified gravity theories have received increased attention in the last years in connection with the dark energy problem, specifically the $f(R)$ theories  \cite{odintsov1}, \cite{odintsov2}, \cite{odintsov3}, \cite{carroll1}, \cite{carroll2}, \cite{dobado}, \cite{luca1}, \cite{luca2}, \cite{starobinsky1}, \cite{starobinsky2}, \cite{wayne1}, \cite{wayne2}. The modified Gauss Bonnet model has also been considered as gravitational alternative for dark energy in \cite{odintsov4}, \cite{sergei4}. A modified $F(R)$ Horava-Lifshitz  gravity and, a unified description of inflation and late-time acceleration in the context of modified $F(R)$ Horava-Lifshitz gravity, have been considered in \cite{odintsov5}, \cite{odintsov6}, \cite{odintsov7}.

Non-minimal coupling of arbitrary function $f(R)$ with matter Lagrangian (including the kinetic scalar term) has been introduced in \cite{sergei},\cite{allemandi}. Such a model was proposed to describe the dark energy and late-time universe acceleration. When $f(R)$ represents the power law function, such a model (which maybe considered as string-inspired theory) maybe proposed for dynamical resolution of cosmological constant problem as it has been suggested in \cite{sergei1},\cite{sergei2}. In the present work we consider this function $f(R)$ as linear in $R$ but we generalize the model permitting extra $R_{\mu\nu}$ coupling with kinetic-like scalar term. In addition, we keep the free kinetic term for the scalar field.\\
\noindent In the present paper we continue the study of the model \cite{granda1}, in which the scalar-field kinetic term is non-minimally coupled to itself and to the scalar and Ricci curvatures, and in addition we consider the matter contribution. The self coupling function $F(\phi)$, allows to find a wide variety of potentials giving rise to accelerated expansion. The power-law solution is considered, and the expressions for the scalar-field and potential are found. The solution corresponding to the cosmological constant with matter ($\Lambda$CDM) is obtained as an exact solution in the present model, giving in this way a dynamical interpretation to the cosmological constant. In addition to the $\Lambda$CDM, other relevant solutions have been studied. In the next section we review the main equations in a general background and in the flat FRW metric.
In section $3$ we study power-law solutions giving rise to accelerated expansion. In section $4$, the $\Lambda$CDM and other relevant solutions for the dark energy are studied, first in the case of scalar field dominance, and then including the matter contribution. Solutions that may accomplish the current observations have been considered. In the last section we present some conclusions.

\section{Field Equations}

Let us start with the following  action, proposed in \cite{granda1}:

\be\label{eq1}
\begin{aligned}
S=&\int d^{4}x\sqrt{-g}\Big[\frac{1}{16\pi G} R-\frac{1}{2}\partial_{\mu}\phi\partial^{\mu}\phi-\frac{1}{2} \xi R \left(F(\phi)\partial_{\mu}\phi\partial^{\mu}\phi\right) -\\ 
&\frac{1}{2} \eta R_{\mu\nu}\left(F(\phi)\partial^{\mu}\phi\partial^{\nu}\phi\right) - V(\phi)\Big] + S_m.
\end{aligned}
\ee

\noindent where $S_m$ is the matter action described by a fluid with barotropic equation of state (along the text, if it is not specified the term ``matter'' will enclose the ordinary barionic matter and dark matter, which obey the same equation of state). The dimensionality of the coupling constants $\xi$ and $\eta$ depends on the type of function $F(\phi)$. Taking the variation of action \ref{eq1} with respect to the metric, we obtain a general expression of the form 
\be\label{eq2}
R_{\mu\nu}-\frac{1}{2}g_{\mu\nu}R=\kappa^2\left[T_{\mu\nu}^m+T_{\mu\nu}\right]
\ee
where $\kappa^2=8\pi G$, $T_{\mu\nu}^m$ is the usual energy-momentum tensor for the ordinary and matter component, and the tensor $T_{\mu\nu}$ represents the variation of the terms which depend on the scalar field $\phi$ and can be written as
\be\label{eq3}
T_{\mu\nu}=T_{\mu\nu}^{\phi}+T_{\mu\nu}^{\xi}+T_{\mu\nu}^{\eta}
\ee
where $T_{\mu\nu}^{\phi}$, $T_{\mu\nu}^{\xi}$, $T_{\mu\nu}^{\eta}$ correspond to the variations of the minimally coupled terms, the $\xi$ and the $\eta$ couplings, respectively with respect to the metric. Due to the interacting terms these expressions are defined in the Jordan frame and do not correspond to the energy-momentum tensors as defined in the Einstein frame. Therefore, due to this interaction between the scalar field and the curvature, the derived expressions for the density and pressure for the scalar field can be regarded as effective ones. Those variations are given by
\be\label{eq4}
T_{\mu\nu}^{\phi}=\nabla_{\mu}\phi\nabla_{\nu}\phi-\frac{1}{2}g_{\mu\nu}\nabla_{\lambda}\phi\nabla^{\lambda}\phi
-g_{\mu\nu}V(\phi)
\ee
\be\label{eq5}
\begin{aligned}
T_{\mu\nu}^{\xi}=&\xi\Big[\left(R_{\mu\nu}-\frac{1}{2}g_{\mu\nu}R\right)\left(F(\phi)\nabla_{\lambda}\phi\nabla^{\lambda}\phi\right)+g_{\mu\nu}\nabla_{\lambda}\nabla^{\lambda}\left(F(\phi)\nabla_{\gamma}\phi\nabla^{\gamma}\phi\right)\\
&-\frac{1}{2}(\nabla_{\mu}\nabla_{\nu}+\nabla_{\nu}\nabla_{\mu})\left(F(\phi)\nabla_{\lambda}\phi\nabla^{\lambda}\phi\right)+R\left(F(\phi)\nabla_{\mu}\phi\nabla_{\nu}\phi\right)\Big]
\end{aligned}
\ee
\be\label{eq6}
\begin{aligned}
T_{\mu\nu}^{\eta}=&\eta\Big[F(\phi)\left(R_{\mu\lambda}\nabla^{\lambda}\phi\nabla_{\nu}\phi+R_{\nu\lambda}\nabla^{\lambda}\phi\nabla_{\mu}\phi\right)-\frac{1}{2}g_{\mu\nu}R_{\lambda\gamma}\left(F(\phi)\nabla^{\lambda}\phi\nabla^{\gamma}\phi\right)\\
&-\frac{1}{2}\left(\nabla_{\lambda}\nabla_{\mu}\left(F(\phi)\nabla^{\lambda}\phi\nabla_{\nu}\phi\right)+\nabla_{\lambda}\nabla_{\nu}\left(F(\phi)\nabla^{\lambda}\phi\nabla_{\mu}\phi\right)\right)\\
&+\frac{1}{2}\nabla_{\lambda}\nabla^{\lambda}\left(F(\phi)\nabla_{\mu}\phi\nabla_{\nu}\phi\right)+\frac{1}{2}g_{\mu\nu}\nabla_{\lambda}\nabla_{\gamma}\left(F(\phi)\nabla^{\lambda}\phi\nabla^{\gamma}\phi\right)\Big]
\end{aligned}
\ee
Varying with respect to the scalar field gives rise to the equation of motion
\be\label{eq7}
\begin{aligned}
&-\frac{1}{\sqrt{-g}}\partial_{\mu}\left[\sqrt{-g}\left(\xi R F(\phi)\partial^{\mu}\phi+\eta R^{\mu\nu}F(\phi)\partial_{\nu}\phi+\partial^{\mu}\phi\right)\right]+\frac{dV}{d\phi}+\\
&\frac{dF}{d\phi}\left(\xi R\partial_{\mu}\phi\partial^{\mu}\phi+\eta R_{\mu\nu}\partial^{\mu}\phi\partial^{\nu}\phi\right)=0
\end{aligned}
\ee
Assuming the spatially-flat Friedmann-Robertson-Walker (FRW) metric,
\be\label{eq8}
ds^2=-dt^2+a(t)^2\left(dr^2+r^2d\Omega^2\right)
\ee
from Eqs. (\ref{eq4}-\ref{eq6}) and using the metric (\ref{eq8}) we can write the $(00)$ and $(11)$ components of the Eq. (\ref{eq2})  (with the Hubble parameter $H$, and for homogeneous time-depending scalar field) as follows
\be\label{eq9}
H^2=\frac{\kappa^2}{3}\rho_{tot}
\ee
and 
\be\label{eq10}
-2\dot{H}-3H^2=\kappa^2 p_{tot},
\ee
with $\rho_{tot}$ given by
\be\label{eq9a}
\begin{aligned}
\rho_{tot}=&\Big[\frac{1}{2}\dot{\phi}^2+V(\phi)+9\xi H^2F(\phi)\dot{\phi}^2+3(2\xi+\eta)\dot{H}F(\phi)\dot{\phi}^2\\
&-3(2\xi+\eta)H F(\phi)\dot{\phi}\hspace{0.1 cm}\ddot{\phi}-\frac{3}{2}(2\xi+\eta)H \frac{dF}{d\phi}\dot{\phi}^3\Big]+\rho_m
\end{aligned}
\ee
and $p_{tot}$ given by
\be\label{eq10a}
\begin{aligned}
p_{tot}=&\Big[\frac{1}{2}\dot{\phi}^2-V(\phi)+3(\xi+\eta)H^2F(\phi)\dot{\phi}^2+2(\xi+\eta)\dot{H}F(\phi)\dot{\phi}^2\\
&+4(\xi+\eta)H F(\phi)\dot{\phi}\hspace{0.1 cm}\ddot{\phi}+2(\xi+\eta)H\frac{dF}{d\phi}\dot{\phi}^3\\
&+(2\xi+\eta)\left(F(\phi)\ddot{\phi}^2+F(\phi)\dot{\phi}\hspace{0.1 cm}\dddot{\phi}+\frac{5}{2}\frac{dF}{d\phi}\dot{\phi}^2\ddot{\phi}+\frac{1}{2}\frac{d^2F}{d\phi^2}\dot{\phi}^4\right)\Big]
\end{aligned}
\ee
here we have assumed that the matter contribution corresponds to presureless perfect fluid (i.e. $p_m=0$). The equation of motion for the scalar field (\ref{eq7}) takes the form
\be\label{eq11}
\begin{aligned}
&\ddot{\phi}+3H\dot{\phi}+\frac{dV}{d\phi}+3(2\xi+\eta)\ddot{H}F(\phi)\dot{\phi}+
3(14\xi+5\eta)H\dot{H}F(\phi)\dot{\phi}\\
&+\frac{3}{2}(2\xi+\eta)\dot{H}\left(2F(\phi)\ddot{\phi}+\frac{dF}{d\phi}\dot{\phi}^2\right)+
\frac{3}{2}(4\xi+\eta)H^2\left(2F(\phi)\ddot{\phi}+\frac{dF}{d\phi}\dot{\phi}^2\right)\\
&+9(4\xi+\eta)H^3F(\phi)\dot{\phi}=0
\end{aligned}
\ee
where ``dot'' denotes derivative with respect to cosmic time. The first three terms in Eq. (\ref{eq11}) correspond to the minimally coupled field. In what follows we study the cosmological consequences of these equations under some conditions that simplify the search for solutions, but nevertheless show the richness of the cosmological dynamics of the present model.\\
\noindent First note that the Eqs. (\ref{eq9}-\ref{eq11}) significantly simplify under the restriction on $\xi$ and $\eta$ given by 
\be\label{eq1a}
\eta+2\xi=0
\ee
This restriction is equivalent to a coupling of the kinetic term to the Einstein tensor $G_{\mu\nu}$ (see \cite{capozziello1}, \cite{capozziello2}), as can be seen from expression (\ref{eq1}). In this case the field equations (\ref{eq9}-\ref{eq11}) contain only second derivatives of the metric and the scalar field. The modified Friedmann equations (\ref{eq9}) and (\ref{eq10}) take the form
\be\label{eq12}
H^2=\frac{\kappa^2}{3}\left(\frac{1}{2}\dot{\phi}^2+V(\phi)+9\xi H^2F(\phi)\dot{\phi}^2+\rho_m\right)
\ee
and
\be\label{eq13}
-2\dot{H}-3H^2=\kappa^2\left[\frac{1}{2}\dot{\phi}^2-V(\phi)-\xi\left(3H^2+2\dot{H}\right)F(\phi)\dot{\phi}^2-2\xi H\left(2F(\phi)\dot{\phi}\hspace{0.1 cm}\ddot{\phi}+\frac{dF}{d\phi}\dot{\phi}^3\right)\right]
\ee
and the equation of motion reduces to
\be\label{eq14}
%\begin{aligned}
\ddot{\phi}+3H\dot{\phi}+\frac{dV}{d\phi}+3\xi H^2\left(2F(\phi)\ddot{\phi}+\frac{dF}{d\phi}\dot{\phi}^2\right)
+6\xi H\left(3H^2+2\dot{H}\right)F(\phi)\dot{\phi}=0
%\end{aligned}
\ee
In what follows we will study cosmological solutions to Eqs. (\ref{eq12}) and (\ref{eq14}), giving rise to accelerated expansion and that satisfy the demands of the actual observational data.\\

%%%%%%%%%%%%%%%%%%%%%%%%%%%%%%%%%%%%%%%%%%%%%%%%%%%%%%%%%%%%%%
%%%%%%%%%%%%%%%%%%%%%%%%%%%%%%%%%%%%%%%%%%%%%%%%%%%%%%%%%%%%%%
\section{Power-law solutions}

Let us consider the model (\ref{eq1}) in the case of scalar field dominance, without matter contribution.
Here we study solutions for the coupling function $F(\phi)$ and the potential $V(\phi)$, which give rise to power-law expansion.\\

\noindent{\bf A. Power-law solutions without potential $V(\phi)=0$}\\

First we consider the model without potential, and analyze the role of the coupling function $F(\phi)$ in the power-law expansion scenario. From Eq. (\ref{eq12}) it follows
\be\label{eq14a}
F(\phi)\dot{\phi}^2=\frac{1}{3\xi\kappa^2}-\frac{\dot{\phi}^2}{18\xi H^2}
\ee
Making $V=0$ in Eq. (\ref{eq14}), multiplying by $\dot{\phi}$ and defining the new variable $\dot{\phi}^2=\psi$, the Eq. (\ref{eq14}) reduces to first order differential equation with respect to both variable $\psi$ and $H$.
\be\label{eq14b}
H\frac{d\psi}{dt}+\left(6H^2-\dot{H}\right)\psi+\frac{6H^2}{\kappa^2}\left(3H^2+2\dot{H}\right)=0
\ee
considering the power-law behavior $H=s/t$, the Eq. (\ref{eq14b}) becomes
\be\label{eq15a}
t^3\frac{d\psi}{dt}+\left(6s+1\right)t^2\psi+\left(3s-2\right)\frac{6s^2}{\kappa^2}=0
\ee
which has the general solution
\be\label{eq16b}
\psi(t)=\dot{\phi}^2=-\frac{6s^2(3s-2)}{(6s-1)\kappa^2t^2}+\frac{\gamma}{t^{1+6s}}
\ee
where $\gamma$ is the integration constant with the appropriate dimension. This equation gives the following solution for the scalar field
\be\label{eq17a}
\phi(t)=\frac{2}{6s-1}\left[\sqrt{\gamma t^{1-6s}-\lambda}+\sqrt{\lambda}\arcsin\left(\sqrt{\frac{\lambda}{\gamma}}t^{3s-1/2}\right)\right],\,\,\,\,\, \lambda=\frac{6s^2(3s-2)}{(6s-1)\kappa^2}
\ee
and the time dependence of the coupling function is obtained as
\be\label{eq18a}
F(t)=\frac{(6s-1)t^{6s+3}}{3\xi\left(\gamma(6s-1)\kappa^2t^2-6s^2(3s-2)t^{6s+1}\right)}
\ee
According to Eq. (\ref{eq17a}) it should be $\lambda>0$, which is accomplished for $s>2/3$ to get real scalar function $\phi(t)$.
In the special case of $s=2/3$ ($\lambda=0$), corresponding to presureless matter dominance, the expression for the scalar field simplifies to $\phi=2/3\gamma^{1/2} t^{-3/2}$, and the explicit dependence of the coupling function on the scalar field becomes
\be\label{eq19a}
F(\phi)=\frac{8}{81\xi \kappa^2}\left(\frac{2\gamma^2}{3}\right)^{1/3}\frac{1}{\phi^{3+1/3}}
\ee
If we take the integration constant $\gamma=0$ in Eq. (\ref{eq16b}), then in order to be consistent with $\psi>0$, $s$ should be in the interval $1/6<s<2/3$. In this case integrating in Eq. (\ref{eq16b}) gives the following solution for the scalar field
\be\label{eq20a}
\phi=\phi_0+\phi_1\log\frac{t}{t_0},\,\,\,\,\,\,\, \phi_1=\sqrt{\frac{6s^2\left(2-3s\right)}{\left(6s-1\right)\kappa^2}}
\ee
with $s$ in the interval $1/6<s<2/3$. Note that the particular value $s=1/2$ corresponding to radiation dominance is an
allowed solution. The coupling function in this case is obtained by making $\gamma=0$, $s=1/2$ and replacing $t$ from Eq. (\ref{eq20a}) in  Eq. (\ref{eq18a}), giving the result (assuming $\phi_0=0$ at $t_0=1$)
\be\label{eq20b}
F(\phi)=\frac{8}{9\xi}\exp\left(\frac{8\kappa}{\sqrt{6}}\phi\right)
\ee
In any case, the solution (\ref{eq16b}) is more appropriate for describing the cosmological dynamics at early times, when the radiation or matter content are dominating.\\
%%%%%%%%%%%%%%%%%%%%%%%%%%%%%%%%%%%%%%%%%%%%%%%%%%%%%%%%%%%%%%%%%%%%%%%%%%%%%%%%%%%%%%%%%%%%
%%%%%%%%%%%%%%%%%%%%%%%%%%%%%%%%%%%%%%%%%%%%%%%%%%%%%%%%%%%%%%%%%%%%%%%%%%%%%%%%%%%%%%%%%%%%
\noindent{\bf B. Power-law solutions with potential $V(\phi)\neq0$}\\

Multiplying the Eq. (\ref{eq14}) by $\dot{\phi}$, defining the function $\psi$ for $\dot{\phi}^2$, and replacing the product $F(\phi)\dot{\phi}^2$ from Eq. (\ref{eq12}),
the Eq. (\ref{eq14}) reduces to first order equation with respect to the variables $\psi$, $H$ and $V$, and can be written as
\be\label{eq21a}
H\frac{d\psi}{dt}+\left(6H^2-\dot{H}\right)\psi+2H\frac{dV}{dt}-2\left(3H^2+\dot{H}\right)V+6\frac{H^2}{\kappa^2}\left(3H^2+2\dot{H}\right)=0
\ee
Note that in Eq. (\ref{eq21a}) the functions $\theta(x)$ and $V(x)$ are not coupled, and we can use this fact in order to simplify the Eq. (\ref{eq21a}), by imposing a restriction on one of these functions (in other words, they may satisfy separate equations by imposing appropriate restrictions on one of the functions). Using this fact, we can limit the model to the class of potentials that satisfy the restriction
\be\label{eq21b}
H\frac{dV}{dt}-\left(3H^2+\dot{H}\right)V+3\frac{H^2}{\kappa^2}\left(3H^2+2\dot{H}\right)=0
\ee
then, the equation for the field $\psi$ significantly simplifies, but still giving an interesting dynamics as can be seen bellow 
\be\label{eq21c}
H\frac{d\psi}{dt}+\left(6H^2-\dot{H}\right)\psi=0
\ee
Let us assume the solution $a(t)\propto t^p$ and replace in Eqs. (\ref{eq21b}, \ref{eq21c}). Is is easy to check that a particular solution to  Eq. (\ref{eq21b}) is given by
\be\label{eq22f}
V(t)=\frac{3p^2\left(3p-2\right)}{\kappa^2\left(3p+1\right)}\frac{1}{t^2}+ C t^{3p-1}
\ee
where $C$ is an integration constant, and the solution to Eq. (\ref{eq21c}) is 
\be\label{eq22g}
\psi=\psi_0\left(\frac{t}{t_0}\right)^{-(6p+1)}
\ee
where $\psi_0$ is the integration constant. The scalar field from Eq. (\ref{eq22g}) is given by
\be\label{eq22ga}
\phi=\phi_0\left(\frac{t}{t_0}\right)^{-3p+1/2},\,\,\,\,\,\,\ \phi_0=\frac{2\sqrt{\psi_0}t_0}{6p-1}
\ee
with these solutions the scalar potential (\ref{eq22f}), and the coupling function from (\ref{eq12}) can be written explicitly in terms of the scalar field as follows
\be\label{eq22h}
V(\phi)=\frac{3p^2\left(3p-2\right)}{\left(3p+1\right)\kappa^2t_0^2}\left(\frac{\phi}{\phi_0}\right)^{\frac{4}{6p-1}}
\ee
and 
\be\label{eq22i}
F(\phi)=\left[\left(\xi\kappa^2\psi_0\right)\left(3p+1\right)\right]^{-1}\left(\frac{\phi_0}{\phi}\right)^{\frac{2(6p+1)}{6p-1}}-\frac{t_0^2}{18\xi p^2}\left(\frac{\phi_0}{\phi}\right)^{\frac{4}{6p-1}}
\ee
which give rise to accelerated expansion for $p>1$.\\ 

%%%%%%%%%%%%%%%%%%%%%%%%%%%%%%%%%%%%%%%%%%%%%%%%%%%%%%%%%%%%%%%%%%%%%%%%%%%%%%%%%%%%%%%%%%%%%%%%%%%%%%%%%
%%%%%%%%%%%%%%%%%%%%%%%%%%%%%%%%%%%%%%%%%%%%%%%%%%%%%%%%%%%%%%%%%%%%%%%%%%%%%%%%%%%%%%%%%%%%%%%%%%%%%%%%%
\section{Cosmological solutions with dynamical equation of state}
Here we study the full model with the restriction (\ref{eq1a}), and will use the freedom in choosing the coupling function in order to obtain a cosmological dynamics closer to the one expected from observations. 
\subsection{Scalar field dominance\label{A}}
Below we study various known proposals for dark energy, including the cosmological constant and matter (named cold dark matter) $\Lambda$CDM, varying cosmological constant $VCC$ and other models.
First we consider the case of scalar field dominance (i.e. $\rho_m=0$). Defining the variables $x$ for $\log a$ and $\theta$ for $\phi'^2$ (``''' denotes the derivative with respect to $x$), the Eq. (\ref{eq14}) can be written as (after multiplying by $\dot{\phi}$)
\be\label{eq48}
\frac{1}{2}\frac{d}{dx}\left(H^2\theta\right)+3H^2\theta+\frac{dV}{dx}+9\xi H^2\frac{dH^2}{dx}F\theta+3\xi H^4\frac{d}{dx}(F\theta)+18\xi H^4F\theta=0
\ee
From Eq. (\ref{eq12}), changing to the variable $x$, we can write the product $F\phi'^2=F\theta$ as following 
\be\label{eq49}
F\theta=\frac{1}{3\xi\kappa^2 H^2}-\frac{\theta}{18\xi H^2}-\frac{V}{9\xi H^4}
\ee
taking the derivative of Eq. (\ref{eq49}) and replacing $F\theta$ and $d(F\theta)/dx$ into Eq. (\ref{eq48}), we arrive at the following equation involving $\theta$, $H$ and $V$
\be\label{eq50}
\begin{aligned}
&2H^4\frac{d\theta}{dx}+H^2\left(12H^2+\frac{dH^2}{dx}\right)\theta+4H^2\frac{dV}{dx}-2\left(6H^2+\frac{dH^2}{dx}\right)V\\
&+12\frac{H^2}{\kappa^2}\left(3H^2+\frac{dH^2}{dx}\right)=0
\end{aligned}
\ee
In this manner, we obtain a first order differential equation for the functions $\theta$, $H$ and $V$. In what follows we will introduce the scaled Hubble parameter $\tilde{H}^2=H^2/H_0^2$ and the dimensionless scalar potential defined as $\tilde{V}=\kappa^2 V/H_0^2$. Note that as in the above studied case, in Eq. (\ref{eq50}) the functions $\theta(x)$ and $V(x)$ are separated (in the sense that are not coupled), so we may proceed as in previous case by imposing a restriction on the scalar field potential $V(x)$, given by the equation 
\be\label{eq51}
2\tilde{H}^2\frac{d\tilde{V}}{dx}-\left(6\tilde{H}^2+\frac{d\tilde{H}^2}{dx}\right)\tilde{V}+6\tilde{H}^2\left(3\tilde{H}^2+\frac{d\tilde{H}^2}{dx}\right)=0
\ee
which simplifies the Eq. (\ref{eq50}):
\be\label{eq52}
2\tilde{H}^2\frac{d\theta}{dx}+\left(12\tilde{H}^2+\frac{d\tilde{H}^2}{dx}\right)\theta=0
\ee
where we have used the new defined scaled Hubble parameter $\tilde{H}$ and potential $\tilde{V}$. 
\noindent{\bf $\Lambda$CDM solution}\\

In order to consistently solve the Eqs. (\ref{eq51}) and (\ref{eq52}), we propose the following expression for the  scaled Hubble parameter $\tilde{H}^2$
\be\label{eq52a}
\tilde{H}^2=\Omega_{m0}e^{-3x}+\Omega_{\Lambda} ,\,\,\,\,\, \Omega_{m0}>0 ,\,\,\,\,\,\, \Omega_{\Lambda}>0
\ee
where $\Omega_{m0}$ can be assumed to be the matter density parameter (this assumption is supported by constraints imposed by different observational datasets), and $\Omega_{\Lambda}$ can be related with the cosmological constant. This is the known $\Lambda$CDM model (note that in this case the solution with cosmological constant is obtained as a result of the dynamics of the present model). It should be stressed the known form of the expansion factor, which from Eq. (\ref{eq52a}) is given by 
\be\label{eq52f}
a(t)=\left(\frac{\Omega_{m0}}{\Omega_{\Lambda}}\right)^{1/3}\left[\sinh\left(\frac{3}{2}\sqrt{\Omega_{\Lambda}}H_0t\right)\right]^{2/3}.
\ee
This solution describes a matter dominated universe at early times ($a(t)\propto t^{2/3}$), and a cosmological constant dominated phase at late times ($a(t)\propto \exp\left(3/2\sqrt{\Omega_{\Lambda}}H_0t\right)$). This model as is well known, is in excellent agreement with different sets of observations.\\
The constants $\Omega_{m0}$ and $\Omega_{\Lambda}$ in (\ref{eq52a}) satisfy the flatness condition at $x=0$ 
\be
\Omega_{m0}+\Omega_{\Lambda}=1
\ee
Replacing  $\tilde{H}^2$ in (\ref{eq51}), after integration we obtain the dimensionless scalar field potential
\be\label{eq52b}
\tilde{V}(x)=6\Omega_{\Lambda}+\frac{6\Omega_{\Lambda}^2 e^{3x}}{\Omega_{m0}}+C e^{3x}\left(\Omega_{\Lambda}+\Omega_{m0}e^{-3x}\right)^{1/2}.
\ee
And replacing $\tilde{H}^2$ in (\ref{eq52}) we get the following expression for $\theta$
\be\label{eq52c}
\theta(x)=\phi'^2=\frac{\theta_0e^{-9x/2}}{\left(\Omega_{\Lambda} e^{3x}+\Omega_{m0}\right)^{1/2}},
\ee
where $\theta_0$ is the integration constant. Integrating the square root of this last equation, we obtain the scalar field as (considering the $(-)$ sign root)
\be\label{eq52d}
\phi(x)=\frac{4\theta_0^{1/2}}{9\Omega_{m0}}\left(\Omega_{\Lambda}+\Omega_{m0}e^{-3x}\right)^{3/4}
\ee
Finally, the coupling function $F$ is found by replacing the Eqs. (\ref{eq52a}), (\ref{eq52b}) and (\ref{eq52c}) in (\ref{eq49}). According to the expressions (\ref{eq52d}) and (\ref{eq52b}), in this case is possible to solve explicitly the potential in terms of the scalar field, giving the result
\be\label{eq52e}
\tilde{V}=\frac{6\Omega_{\Lambda}\gamma \phi^{4/3}+C\Omega_{m0}\gamma^{1/2}\phi^{2/3}}{\gamma \phi^{4/3}-\Omega_{\Lambda}},\,\,\, \gamma=(\frac{3}{2})^{8/3}\Omega_{m0}^{4/3}\theta_0^{-2/3}
\ee
where we assume the restriction $\gamma\phi^{4/3}>\Omega_{\Lambda}$.\\
%%%%%%%%%%%%%%%%%%%%%%%%%%%%%%%%%%%%%%%%%%%%%%%%%%%%%%%%%%%%%%%%%%%%%%%%%%%%%%%%%%%%%
\noindent{\bf  VCC-type solution}\\

We will consider the following solution, which generalizes the $H$ power-law model in the context of the variable cosmological constant theories (VCC) \cite{yinze}, \cite{marek}.
\be\label{eq52v}
\tilde{H}^2=\left(A e^{-\alpha x}+B\right)^{\beta}
\ee
where $A$, $\alpha$, $B$ and $\beta$ are constants that can be fixed by the expected behavior of the main cosmological parameters (The equation of state parameter (EoS) $\text{w}$, the deceleration parameter $q$, etc.), or can be constrained using the different observational datasets. Replacing (\ref{eq52v}) in (\ref{eq51}), we obtain the scalar field potential of the form
\be\label{eq52v1}
\begin{aligned}
&\tilde{V}(x)=C e^{3x}\left(A e^{-\alpha x}+B\right)^{\beta/2}+\frac{6A^{\beta/2-1}e^{-\alpha\beta x}\left(A+Be^{\alpha x}\right)^{\beta/2}}{(\alpha\beta+6)(\alpha\beta-2\alpha+6)}\times \\
&\Big[3B(\alpha\beta+6)e^{\alpha x}\hspace{0.2cm} _{2}F_1\left[1-\frac{\beta}{2},1-\frac{3}{\alpha}-\frac{\beta}{2},2-\frac{3}{\alpha}-\frac{\beta}{2},-\frac{Be^{\alpha x}}{A}\right]\\
&\left(6\alpha+3\alpha\beta-2\alpha^2\beta+\alpha^2\beta^2-18\right)\hspace{0.2cm} _{2}F_1\left[1-\frac{\beta}{2},-\frac{3}{\alpha}-\frac{\beta}{2},1-\frac{3}{\alpha}-\frac{\beta}{2},-\frac{Be^{\alpha x}}{A}\right]\Big]
\end{aligned}
\ee
where $ _{2}F_1$ is the hypergeometric function. Integrating the Eq. (\ref{eq52}) we obtain the expression for $\theta$
\be\label{eq52v2}
\theta(x)=\theta_0 e^{(\alpha\beta-12)x/2}\left(A+Be^{\alpha x}\right)^{-\beta/2}
\ee
where $C$ and $\theta_0$ are integration constants. After integrating the square root of (\ref{eq52v2}), gives the scalar field
\be\label{eq52v3}
\phi(x)=\frac{4\theta_0^{1/2}A^{-\beta/4}}{\alpha\beta-12}e^{(\alpha\beta-12)x/4}\hspace{0.2cm} _{2}F_1\left[-\frac{3}{\alpha}+\frac{\beta}{4},\frac{\beta}{4},1-\frac{3}{\alpha}+\frac{\beta}{4},-\frac{Be^{\alpha x}}{A}\right]
\ee
Finally, replacing $\tilde{H}^2$, $\tilde{V}$ and $\theta$ from (\ref{eq52v})-(\ref{eq52v2}) in the generalized Friedmann equation (\ref{eq49}), we obtain the expression for the coupling function $F$, thus completing the solution to the
Eqs. (\ref{eq49})  and (\ref{eq50}) under the restriction (\ref{eq51}).
Here the parameters $A$ and $B$ must satisfy the condition $A+B=1$ at $x=0$, as follows from (\ref{eq52v}). If we take $A=\Omega_{m0}$, $\alpha=3(1-\frac{n}{2})$ and $\beta=2/(2-n)$ in (\ref{eq52v}), then we arrive at the solution to a variable cosmological model of the $H$ power-law type with matter-dark energy interaction \cite{yinze}.\\

\noindent{\bf  FRW with 2D topological defects}\\

An analytical expression for the potential and the scalar field can be found for the Hubble function of the form
\be\label{eq52n}
\tilde{H}^2=\Omega_{m0}e^{-3x}+\Omega_{T0}e^{-x},
\ee
which describes the FRW cosmology with 2D topological defects, with the EoS of the form $p_X=-2/3\rho_X$ \cite{marek,marek1}.
Replacing in Eq. (\ref{eq52}) we obtain the following expression for the potential
\be\label{eq52n1}
V(x)=C e^{3x/2}\left(\Omega_{m0}+\Omega_{T0}e^{2x}\right)+\frac{12}{5}\Omega_{T0}\left(e^{-2x}+\frac{\Omega_{T0}}{\Omega_{m0}}\right)\hspace{0.2cm} _{2}F_1\left[-\frac{5}{4},\frac{1}{2},-\frac{1}{4},-\frac{\Omega_{T0}e^{2x}}{\Omega_{m0}}\right]
\ee
and the function $\theta$ is given by
\be\label{eq52n2}
\theta(x)=\sqrt{2}\theta_0 e^{-9x/2}\left(\Omega_{m0}+\Omega_{T0} e^{2x}\right)^{1/2}
\ee
However, when a fixed value for $\Omega_{m0}$ is considered, then while $\Omega_{m0}<0.22$ the model with 2D topological defects gives an appropriate description of the EoS, according to the current observations (i.e. this model favors a low density universe \cite{marek1}).\\

\noindent{\bf  FRW with phantom dark energy}\\

The following expression describes the FRW solution with phantom dark energy, with EoS for the phantom sector of the form $p_X=\text{w}_X\rho_X$ \cite{marek}
\be\label{eq52p}
\tilde{H}^2=\Omega_{m0}(1+z)^{3}+\Omega_{P0}(1+z)^{3(1+\text{w}_X)}
\ee
where $\text{w}_X<-1$ and the parameters $\Omega_{m0}$ and $\Omega_{P0}$ satisfy the restriction at the present epoch $\Omega_{m0}+\Omega_{P0}=1$. Turning to the $x$-variable and replacing this proposal in Eq. (\ref{eq51}), we get the potential
\be\label{eq52p1}
\begin{aligned}
\tilde{V}(x)=&Ce^{3x/2}\sqrt{\Omega_{m0}+\Omega_{P0}e^{-3\text{w}_X x}}-\\
&\frac{6\Omega_{P0}\text{w}_X}{3+\text{w}_X}e^{-3(1+\text{w}_X)x}\sqrt{1+\frac{\Omega_{m0}e^{3\text{w}_X x}}{\Omega_{P0}}}\hspace{0.2cm} _{2}F_1\left[\frac{1}{2},-\frac{3+\text{w}_X}{2\text{w}_X},\frac{-3+\text{w}_X}{2\text{w}_X},-\frac{\Omega_{m0}e^{3\text{w}_X x}}{\Omega_{P0}}\right]
\end{aligned}
\ee
where $C$ is the integration constant. From Eq. (\ref{eq52}) it follows the expression for the function $\theta(x)$
\be\label{eq52p2}
\theta(x)=\theta_0\frac{e^{-9/2x}}{\left(\Omega_{m0}+\Omega_{P0}e^{-3\text{w}_X x}\right)^{1/2}}
\ee
replacing $\tilde{H}^2$, $\tilde{V}$ and $\theta$ from (\ref{eq52p})-(\ref{eq52p2}) in (\ref{eq49}), we obtain the expression for the coupling function $F$, which allows to completely solve the generalized Friedmann 
Eq. (\ref{eq49}) and equation of motion (\ref{eq50}), under the restriction (\ref{eq51}). In this case it is interesting to analyze the EoS parameter, which as expected presents quintom behavior \cite{yicai}. To this end, we write the effective equation of state parameter in terms of the redshift
\be\label{eq52p3}
\text{w}_{eff}(z)=-1+\frac{1}{3}\frac{(1+z)}{\tilde{H}^2}\frac{d(\tilde{H}^2)}{dz}
\ee 
\noindent replacing $\tilde{H}^2$ from Eq. (\ref{eq52p}) we obtain
\be\label{eq52p4}
\text{w}_{eff}(z)=-1+\frac{(1+z)\left[3\Omega_{m0}(1+z)^2+3(1-\Omega_{m0})(1+\text{w}_X)(1+z)^{3(1+\text{w}_X)-1}\right]}{3\left[\Omega_{m0}(1+z)^3+(1-\Omega_{m0})(1+z)^{3(1+\text{w}_X)}\right]}
\ee
where we used $\Omega_{P0}=1-\Omega_{m0}$. Evaluating this expression at the present epoch ($z=0$) gives $\text{w}_{eff0}=\text{w}_X-\Omega_{m0}\text{w}_X$, showing that this model crosses the phantom barrier and presents quintom behavior for $\Omega_{m0}<(\left|\text{w}_X\right|-1)/\left|\text{w}_X\right|$. In Fig. 1 we plot $\text{w}_{eff}(z)$ for two representative values of $\text{w}_X$, taking $\Omega_{m0}=0.27$.
\begin{center}
\includegraphics [scale=0.7]{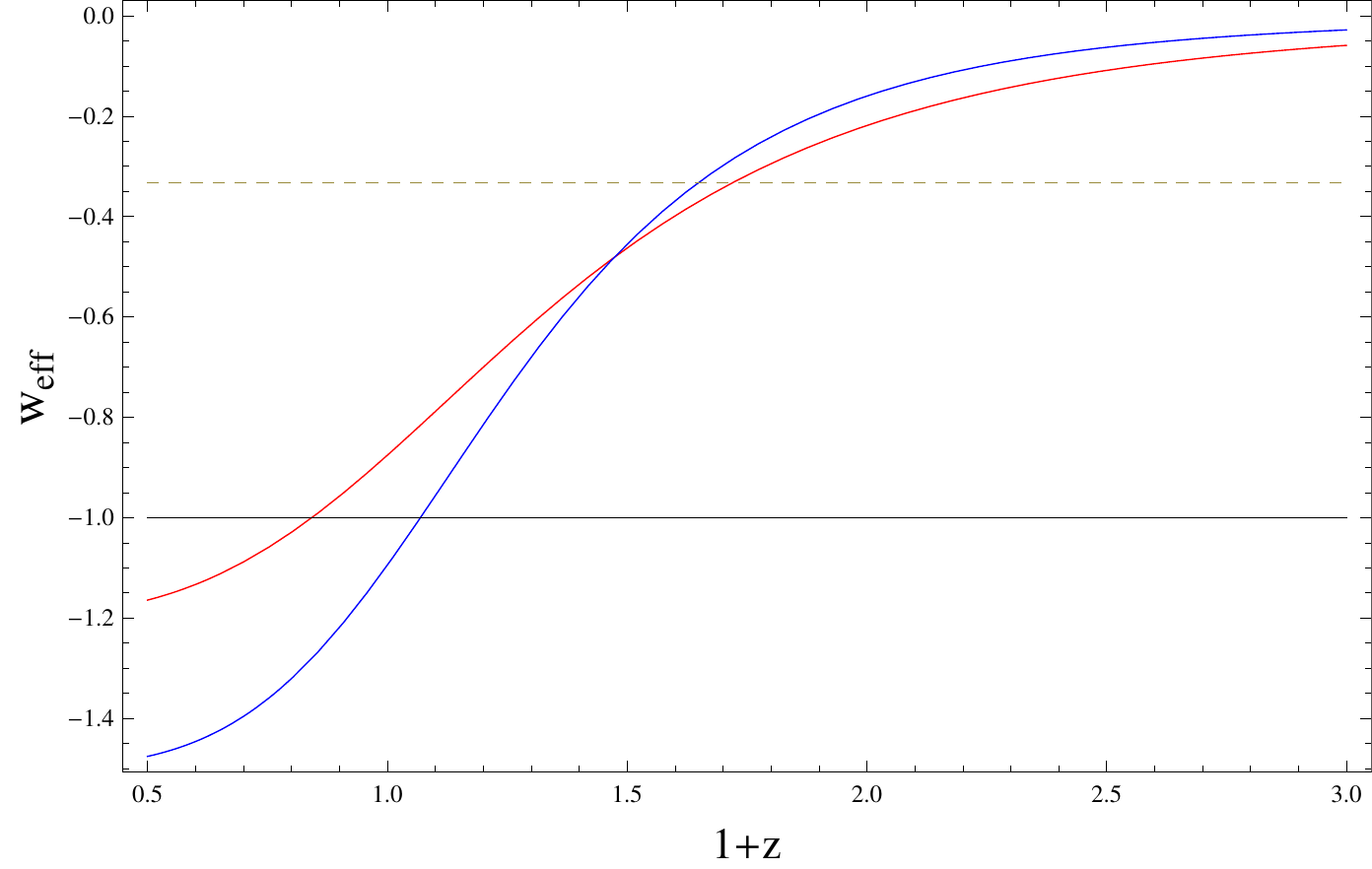}
\end{center}
\begin{center}
{Fig. 1 \it The effective equation of state} $\text{w}_{eff}$ {\it versus redshift, for two values of} $\text{w}_X$ {\it: red-}($\text{w}_X=-1.2$){\it, blue-}($\text{w}_X=-1.5$). 
{\it The continue line }$\text{w}_{eff}=-1$ { \it corresponds to the cosmological constant. The dashed line} ($\text{w}_{eff}=-1/3$) {\it intersects the curves at $z_T\sim 0.65-0.72$. Note that the smaller the} $\left|\text{w}_X\right|${\it is, the earlier the redshift transition occurs}
\end{center}
%%%%%%%%%%%%%%%%%%%%%%%%%%%%%%%%%%%%%%%%%%%%%%%%%%%%%%%%%%%%%%%%%%%%%%%%%%%%%%%%%%%%%%%%%%%%%%%%%%%%%%%%%%%%%%%%%%%%%%%%%%
\subsection{Solutions with both scalar field and matter contributions \label{B}}

Let us now focus in the solutions with the contribution of the matter term. Here we will consider the $\Lambda$CDM and a more general model of dark energy.
From Eq. (\ref{eq12}) and changing to the variable $x$, we can write the product $H^2F\theta$ as follows
\be\label{eq53}
H^2F\theta=\frac{1}{3\xi\kappa^2}-\frac{\theta}{18\xi}-\frac{V}{9\xi H^2}-\frac{\rho_{m0} e^{-3x}}{9\xi H^2}
\ee
where we used the expression for the presureless matter component $\rho_m=\rho_{m0}e^{-3x}$. Taking the derivative with respect to $x$ in Eq. (\ref{eq53}), and replacing $F\theta$ and $d(F\theta)/dx$ into Eq. (\ref{eq48}), we arrive at the following equation involving $\theta$, $H$ and $V$
\be\label{eq54}
\begin{aligned}
&2H^4\frac{d\theta}{dx}+H^2\left(12H^2+\frac{dH^2}{dx}\right)\theta+4H^2\frac{dV}{dx}-2\left(6H^2+\frac{dH^2}{dx}\right)V\\
&+\frac{12}{\kappa^2}H^2\left(3H^2+\frac{dH^2}{dx}\right)-\rho_{m0}\left(3H^2+\frac{dH^2}{dx}\right)e^{-3x}=0
\end{aligned}
\ee
This is a first order differential equation for the functions $\theta$, $H$ and $V$. Starting from Eq. (\ref{eq54}) we will consider different possibilities of cosmological scenarios. In order to integrate the equation (\ref{eq54}), we use again the fact that the functions $\theta$ and $V$ are separated, and thus we may constrain the scalar field potential. It is important to note that we have different possibilities for the restrictions we can impose on the potential.\\
\noindent{\bf First restriction}\\
Here we consider the following restriction on the potential, using the above defined $\tilde{H}$ and $\tilde{V}$
\be\label{eq55}
2\tilde{H}^2\frac{d\tilde{V}}{dx}-\left(6\tilde{H}^2+\frac{d\tilde{H}^2}{dx}\right)\tilde{V}+6\tilde{H}^2\left(3\tilde{H}^2+\frac{d\tilde{H}^2}{dx}\right)-3\Omega_{m0}\left(3\tilde{H}^2+\frac{d\tilde{H}^2}{dx}\right)e^{-3x}=0
\ee
which after the replacement in (\ref{eq54}), simplifies the equation for $\theta$ :
\be\label{eq56}
2\tilde{H}^2\frac{d\theta}{dx}+\left(12\tilde{H}^2+\frac{d\tilde{H}^2}{dx}\right)\theta=0
\ee
where $\Omega_{m0}$ is the matter density parameter defined as $\Omega_{m0}=\kappa^2\rho_{m0}/(3H_0^2)$. Note that this equation for $\theta$ is the same as obtained in (\ref{eq52}).\\

\noindent{\bf $\Lambda$CDM solution}\\

Let us propose again the $\Lambda$CDM solution for $\tilde{H}$ given by (\ref{eq52a}), and evaluate $\tilde{V}$ and $\theta$ from the Eqs. (\ref{eq55}) and (\ref{eq56}).
Replacing  $\tilde{H}^2$ in (\ref{eq55}), after integration we obtain the following expression for the dimensionless scalar field potential
\be\label{eq57}
\tilde{V}(x)=3\Omega_{\Lambda}+ C e^{3x}\left(\Omega_{\Lambda}+\Omega_{m0}e^{-3x}\right)^{1/2},
\ee
with $C$ as an integration constant. From Eq. (\ref{eq56}), it follows the expression for $\theta$
\be\label{eq58}
\theta(x)=\phi'^2=\frac{\theta_0e^{-9x/2}}{\left(\Omega_{\Lambda} e^{3x}+\Omega_{m0}\right)^{1/2}},
\ee
where $\theta_0$ is the integration constant. Integrating the square root of Eq. (\ref{eq57}), we find the same solution for the scalar field obtained in (\ref{eq52d}). The explicit dependence of the potential on the scalar field is obtained from Eqs. (\ref{eq52d}) and (\ref{eq57})   
\be\label{eq59}
\tilde{V}=3\Omega_{\Lambda}+\frac{C\gamma^{1/2}\Omega_{m0}\phi^{2/3}}{\gamma\phi^{4/3}-\Omega_{\Lambda}}.
\ee
From Eqs. (\ref{eq52b}) and (\ref{eq57}) it follows that the dark energy potentials (without and with matter content) describing the $\Lambda$CDM model, increase as the
Universe expands ($x\rightarrow \infty$).\\

\noindent{\bf Phenomenologically the desired solution}\\

In this case we consider an interesting solution for the Hubble parameter, which is usually introduced phenomenologically to describe the dark energy, and successfully  fit the observations (see \cite{yinze})
\be\label{eq60}
\tilde{H}^2=\Omega_{m0} e^{-3x}+ \Omega_{\alpha}e^{-\alpha x}=\Omega_{m0}\left(1+z\right)^3+ \Omega_{\alpha}\left(1+z\right)^{\alpha}
\ee
where $\Omega_{m0}$ and $\Omega_{\alpha}$ are subject to the restriction
\be
\Omega_{m0}+\Omega_{\alpha}=1
\ee
replacing the solution (\ref{eq60}) in Eq. (\ref{eq51}) we obtain the following expression for the scalar field potential 
\be\label{eq61}
\begin{aligned}
&\tilde{V}(x)=Ce^{-\frac{1}{2}(\alpha-3)x}\left(\Omega_{m0}e^{\alpha x}+\Omega_{\alpha} e^{3x}\right)^{1/2}- \frac{3\Omega_{\alpha}(\alpha-3)}{\Omega_{m0}(2\alpha+3)(4\alpha-3)}\left(1+\frac{\Omega_{\alpha}}{\Omega_{m0}}e^{(3-\alpha)x}\right)^{1/2}\times\\
& \Big[2(2\alpha+3)\Omega_{\alpha}\hspace{0.2cm} _{2}F_1\left[\frac{3-4\alpha}{6-2\alpha},\frac{3}{2},\frac{9-6\alpha}{6-2\alpha},-\frac{\Omega_{\alpha}}{\Omega_{m0}}e^{(3-\alpha)x}\right]+\\&
(4\alpha-3)\Omega_{m0} e^{(\alpha-3)x}\hspace{0.2cm} _{2}F_1\left[\frac{2\alpha+3}{2(\alpha-3)},\frac{3}{2},\frac{3-4\alpha}{6-2\alpha},-\frac{\Omega_{\alpha}}{\Omega_{m0}}e^{(3-\alpha)x}\right]\Big]
\end{aligned}
\ee
and replacing (\ref{eq60}) in Eq. (\ref{eq55}) gives the solution for $\theta$ 
\be\label{eq62}
\theta(x)=\theta_0\frac{e^{(\alpha-9)x}}{\sqrt{\Omega_{\alpha}e^{3x}+\Omega_{m0}e^{\alpha x}}}
\ee
which in turns gives the scalar field 
\be\label{eq63}
\phi(x)=\phi_0 e^{-9x/4}\hspace{0.2cm} _{2}F_1\left[\frac{9}{4(\alpha-3)},\frac{1}{4},1+\frac{9}{4(\alpha-3)},-\frac{\Omega_{\alpha}}{\Omega_{m0}}e^{(3-\alpha)x}\right]
\ee
where $\phi_0=4\theta_0^{1/2}/(9\Omega_{m0}^{1/4})$.\\
The parameters $\Omega_{m0}$ and $\alpha$ in (\ref{eq60}), can be constrained by demanding that the current value of the EoS parameter be $\text{w}_0=-1$. To his end, we replace the scaled Hubble function (\ref{eq60}) (given in terms of the redshift) in the expression for $\text{w}_{eff}$ (\ref{eq52p3}), and impose the condition $\text{w}_{eff}(0)=\text{w}_0=-1$. After that,
we obtain the relation
\be\label{eq64}
\alpha=\frac{3\Omega_{m0}}{\Omega_{m0}-1}
\ee
If we take for instance the value $\Omega_{m0}=0.27$, then $\alpha \sim -1.11$. We can also demand that $\text{w}_0=-1.1$, in which case for $\Omega_{m0}=0.3$, $\alpha$ takes the value $\alpha \sim -1.71$ and the model presents quintom behavior. The behavior of the effective EoS parameter, for both cases is shown in Fig. 2
\begin{center}
\includegraphics [scale=0.7]{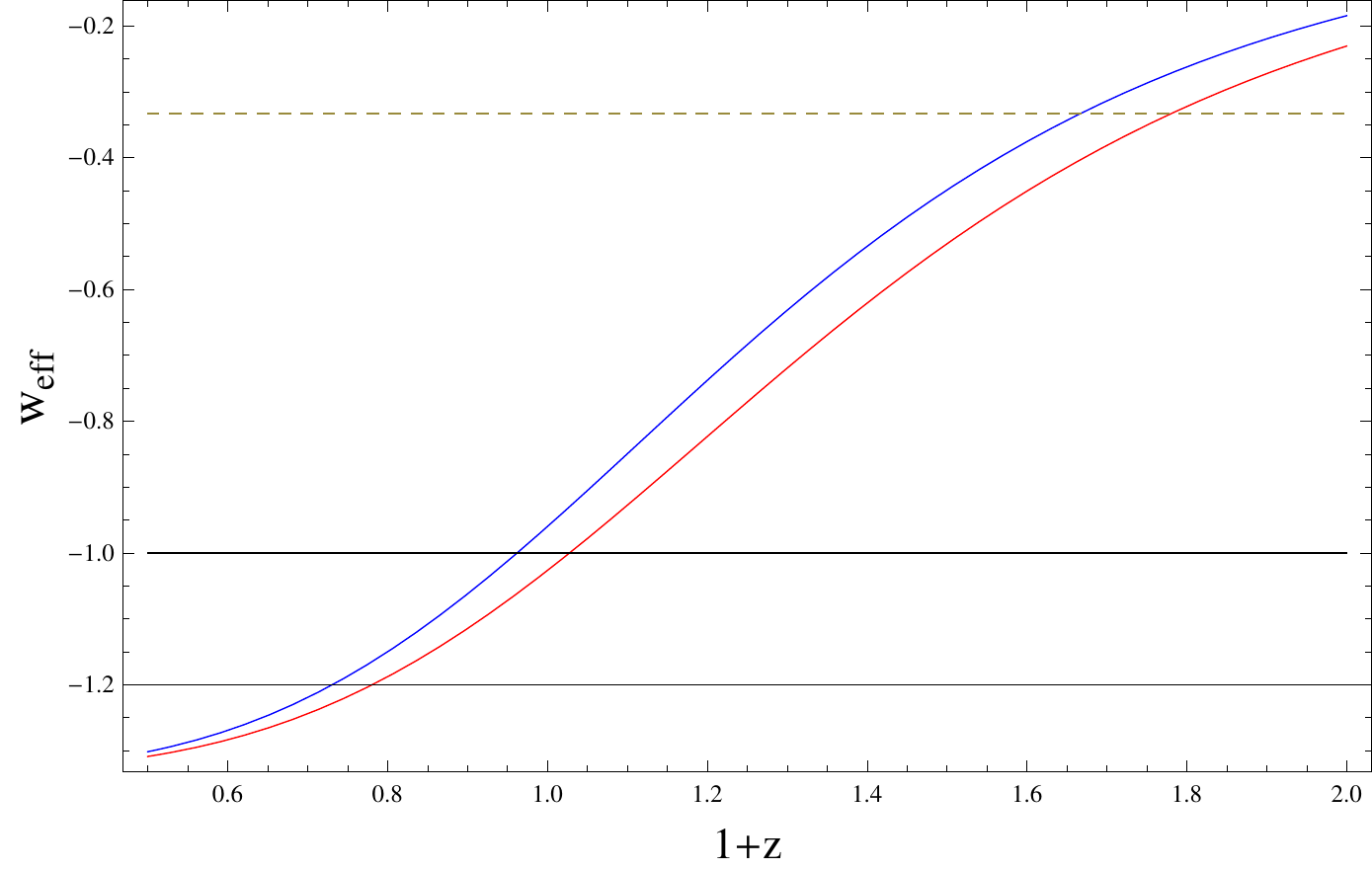}
\end{center}
\begin{center}
{Fig. 2 \it The effective equation of state} $\text{w}_{eff}$ { \it versus redshift, for two sets of parameters: red-($\Omega_{m0}=0.27, \alpha=-1.11$), blue-($\Omega_{m0}=0.3, \alpha=-1.71$). The continue line} $\text{w}_{eff}=-1$ { \it corresponds to the cosmological constant. The dashed line} ($\text{w}_{eff}=-1/3$) {\it intersects the curves at $z_T\sim 0.57-0.68$. Note that the increase in $\Omega_{m0}$ delay the onset of the redshift transition.}
\end{center}
Note that the curves in Fig. 1 and 2 correctly describe the matter dominance at early times, the transition to accelerated expansion at $z\sim 0.57-0.72$ (see \cite{cunha}), and the current accelerated expansion with the equation of state very close to $-1$.\\
Concerning the time variation of the gravitational coupling, and applied to the solutions found here, we can proceed in the same way as in \cite{granda} and
\cite{granda1}. 
The effective gravitational coupling from (\ref{eq12}) can be written as
\be\label{eqgo}
G_{eff}=\frac{G}{1-3\xi \kappa^2F H^2\theta}
\ee
where we used $\kappa^2=8\pi G$ and $\dot{\phi}^2=H^2\phi'^2=H^2\theta$. Then, in terms of $x$ the time variation of the gravitational coupling can be written as (taking into account that $d/dt=Hd/dx$)
\be\label{eqg1}
\frac{\dot{G}_{eff}}{G_{eff}}=\frac{3\xi \kappa^2}{1-3\xi \kappa^2FH^2\theta}\frac{d}{dx}(FH^2\theta)H.
\ee
\noindent Replacing the product $FH^2\theta$ from Eqs. (\ref{eq49}) or  (\ref{eq53}), and evaluating at the present time ($x=0$), the Eq. (\ref{eqg1}) can be written as
\be\label{eqg2}
\frac{\dot{G}_{eff}}{G_{eff}}\Big|_{x=0}=\frac{3f(C,\theta_0)}{1-3g(C,\theta_0)}H_0
\ee
where the main parameters of the model appearing in $\tilde{H}^2$ have been fixed, and the only free parameters present in (\ref{eqg2}) are the integrations constants $C$ and $\theta_0$. This defines the functions $f$ and $g$ through the expressions $f(C,\theta)=\xi \kappa^2FH^2\theta$ and $g=\xi \kappa^2\frac{d}{dx}(FH^2\theta)$, valuated at $x=0$ and at the values of the parameters that have been previously fixed.  
We can meet the constraints on the current value and the time variation of the gravitational coupling \cite{uzan1}, by appropriately defining or constraining the constants $C$ and $\theta_0$. Here we illustrate the important case of the $\Lambda$CDM solution (\ref{eq52a}) with (\ref{eq57}) and (\ref{eq58}). Replacing (\ref{eq52a}), (\ref{eq57}) and (\ref{eq58}) in Eq. (\ref{eq53}), and then in (\ref{eqg1}) (making the needed changes to $\tilde{H}^2$ and $\tilde{V}$ in (\ref{eqg1})), and evaluating at the present time, it is obtained
\be\label{eqg3}
\frac{\dot{G}_{eff}}{G_{eff}}\Big|_{x=0}=\frac{\frac{5}{6}\kappa^2\theta_0-(2+\Omega_{m0})\frac{C}{2}}{1+\frac{\kappa^2\theta_0}{6}\left[1+\frac{10}{3(2+\Omega_{m0})}\right]}H_0
\ee
In this case we can satisfy the requirements of the time variation of the gravitational coupling, by defining the constant $C$ as
$C=\frac{5\kappa^2\theta_0}{3(2+\Omega_{m0})}$, annulling the expression (\ref{eqg3}). And by choosing $\theta_0=H_0^2$, we make the second term in the denominator (and therefore in the denominator of (\ref{eqgo}) which affects the current value of the Newtonian coupling) extremely small, all in complete agreement with the observed constraints.\\

\noindent{\bf Other restrictions on the potential}\\

Besides the above restriction (\ref{eq55}), and due to the extra term in Eq. (\ref{eq54})  related to matter,  we can consider other possible restrictions on the potential $V(\phi)$. Consider for example the same restriction given by Eq. (\ref{eq51}). After the replacement in (\ref{eq54}), the corresponding equation for the function $\theta(x)$ becomes
\be\label{eq65}
2\tilde{H}^4\frac{d\theta}{dx}+\tilde{H}^2\left(12\tilde{H}^2+\frac{d\tilde{H}^2}{dx}\right)\theta-3\Omega_{m0}\left(3\tilde{H}^2+\frac{d\tilde{H}^2}{dx}\right)e^{-3x}=0
\ee
Thus, we keep the same solutions for the Hubble parameter and the potential as obtained in subsection \ref{A}, but according to (\ref{eq65}) the equation for the scalar field becomes more complicated and some solutions can not be expressed analytically, and should be solved numerically.\\
\noindent Other  possible restrictions on the potential are given by
\be\label{eq67}
2\tilde{H}^2\frac{d\tilde{V}}{dx}-\left(6\tilde{H}^2+\frac{d\tilde{H}^2}{dx}\right)\tilde{V}-3\Omega_{m0}\left(3\tilde{H}^2+\frac{d\tilde{H}^2}{dx}\right)e^{-3x}=0
\ee
or
\be\label{eq68}
2\tilde{H}^2\frac{d\tilde{V}}{dx}-\left(6\tilde{H}^2+\frac{d\tilde{H}^2}{dx}\right)\tilde{V}=0, 
\ee

\noindent Hence, starting from the model with the action (\ref{eq1}) we obtained a variety of solutions which not only give the accelerated expansion, but also give the correct transition from decelerating phase to accelerating phase (for transition deceleration-acceleration in f(G)-gravity see \cite{sergeio}), in agreement with the  observational data. The fact that the present model contains the type of solutions as described above in subsections \ref{A} and \ref{B}, indicates that the scalar field with non-minimally kinetic coupled terms, as presented here, could explain the dynamical origin of these (mostly) phenomenological models of dark energy. On the other hand, in order to obtain the above solutions (see section \ref{A}) it was not necessary to introduce the matter term in the action. The introduction of matter term translates into extra term in the Eq. (\ref{eq54}), allowing  for different restrictions on the potential, which may enrich the dynamical possibilities of the present model. \\

\section{Discussion and conclusions}
We have considered a model of scalar field with kinetic terms coupled non-minimally to the scalar field and to the curvature. We found power-law solutions without and with a potential, giving rise to different power-law scenarios which include radiation dominance, matter dominance and also describing accelerated expansion. 
Due to the complexity of the system of equations which describes the dynamics of the model, and thanks to the presence of the coupling function $F(\phi)$, we could impose a restriction on the potential through Eq. (\ref{eq51}) or (\ref{eq55}) (if explicitly matter content is added), which allowed us to find a class of solutions that lead to dynamical effective EoS. We proposed different solutions for the Hubble function, corresponding to known important models of dark energy, like $\Lambda$CDM, variable cosmological constant model VCC, FRW with two dimensional topological defects, FRW with phantom dark energy, and so on. These solutions properly describe the accelerated expansion and the transition from the decelerated to accelerated phase, according to the current astrophysical observations, and also account for the cosmic coincidence (i.e. that guarantee the matter dominance at early times (high redshift), and the dark energy dominance at late times). This means that the present model could explain the dynamical origin of many phenomenologically viable dark energy models.\\
\noindent One relevant fact of this model, is that it contains solutions with EoS that cross the phantom frontier, showing that the present model with non-minimal kinetic coupling is able to describe the quintom behavior.
We also illustrated the time variation of the gravitational coupling for the relevant case of the $\Lambda$CDM solution. But in general we can proceed with the Eq. (\ref{eqg1}), and in order to satisfy the current restrictions on the Newtonian coupling, we can use the freedom in the integrations constants $C$ and $\theta_0$, in order to control the the actual value of the Newton's constant and its time variation. These conditions can be satisfied by imposing the inequalities $f(C,\theta_0)\leq10^{-1}$ and $g(C,\theta_0)\leq 10^{-5}$.\\
\noindent In conclusion, the derivative couplings provide an effective scalar density and pressure which might play an important role in the explanation of the dark energy or
cosmological constant. The present model shows a wide variety of acceptable solutions, and particularly we found solutions for the Hubble parameter, used to describe the actual phenomenology of the dark energy, according to the current observational data.

%\section*{Acknowledgments}
%This work was supported by the SENECA foundation under the PCTRM 2007-2010.

\end{document}